\documentstyle[12pt,epsfig]{article} 
\begin{document} 
\begin{titlepage}
\begin{flushright} 
TAUP-2342-96\\ 
hep-ph/9606284\\ 
June 1996
\end{flushright} 
\vskip 1cm 
\begin{center} 
{\Large \bf Passage of charmed
particles through the mixed phase \\ 
in high-energy heavy-ion collisions
\\ \ \\} 
Benjamin Svetitsky\footnote{E-mail: bqs@julian.tau.ac.il}\\ 
Asher Uziel\footnote{Current address: Scitex Corporation,
P.O.B. 330, 46103 Herzlia B, Israel}\\ 
\ \\ 
{\it School of Physics and Astronomy\\ 
Raymond and Beverly Sackler Faculty\\ 
of Exact Sciences\\ 
Tel Aviv University\\ 
69978 Tel Aviv, Israel}
\end{center} 
\vskip .75cm 
\centerline{PACS numbers: 12.38.Mh, 12.38.Qk, 14.40.Lb, 52.25.Fi}
\vskip .75cm
\noindent 
{\it ABSTRACT.\/} We employ a modified cascade hydrodynamics code to
simulate the phase transition of an expanding quark-gluon plasma and
the passage of a charmed particle through it.
When inside the plasma droplets, the charmed quark experiences drag and
diffusion forces.
When outside the plasma, the quark travels as a $D$ meson and
experiences collisions with pions.
Additional energy transfer takes place when the quark enters or leaves
a droplet.  We find that the transverse momentum of $D$ mesons
provides a rough thermometer of the phase transition.

\end{titlepage}

\section{Introduction}

Charmed particles are promising candidates for probing the early stages
of ultrarelativistic heavy-ion collisions.
Because of their large mass, they can be created only in the initial
hard collisions between the nuclear constituents \cite{Shor}.
Most charmed quarks created in hadronic collisions emerge with
nonrelativistic velocities \cite{charm_production}, and hence in
nuclear collisions they will remain in the collision region for a long
time.
Finally, the passage of charmed quarks through the plasma is a problem
amenable to theoretical analysis \cite{old_drag,bqs_drag,daphne,other_drag}.
Thus the observation of charmed particles, and in particular the
measurement of their transverse momentum distribution, should yield
information about the hot environment present early in the system's
evolution.

Consider a high-energy nuclear collision which leads to creation in the
central rapidity region of both a charmed quark pair and a quark-gluon
plasma.
We adopt Bjorken's scaling hydrodynamics {\em ansatz} to describe the
plasma \cite{pre_Bj,Bj}.
The initial creation of the $c\bar c$ pair takes place in a time on the
order of $1/m_c \simeq 0.1$~fm/$c$.
Within a few tenths of fm/$c$ the quarks will find themselves in an
expanding quark-gluon plasma near local equilibrium
\cite{qgp_equilibration}.
We begin following the progress of each quark at this time,
$\tau=\tau_0$, when the temperature is $T_0$.
At this point, we assume, each quark begins a process of scattering and
diffusion in the plasma which causes a loss of initial momentum and
relaxation towards the (changing) thermal velocity.

The plasma expands and cools according to $T=T_0\cdot
(\tau/\tau_0)^{-1/3}$, until at $\tau=\tau_p$ the phase transition
begins, the temperature having reached $T=T^*$, the transition
temperature.
The ``mixed phase'' forms as follows.
If there is a genuine first-order transition, the plasma supercools
slightly, and then bubbles of the low-temperature phase are nucleated
by fluctuations or by inhomogeneities.
These bubbles grow, meet, and coalesce, and eventually we have a hadron
gas containing mostly pions and also droplets of plasma which are
evaporating.
The phase transition is complete when the last of the droplets
evaporate into hadrons, at $\tau=\tau_h$.
The enormous entropy of the plasma which must be converted to hadrons
dictates that $\tau_h$ is more than ten times $\tau_p$:
The mixed phase is the longest stage in the evolution of the system.

At the beginning of the mixed phase each charmed quark is in plasma,
but it may emerge into a region of hadron gas, and then it might be
reabsorbed into plasma only to emerge again\cite{bqs_mixed}.
Emergence into the hadron gas requires hadronization, the binding of
the charmed quark with a light antiquark to form a meson; reabsorption
involves stripping off the antiquark by collisions with the
plasma.
We will discuss the processes of hadronization and ionization in detail
below; suffice it for now to observe that hadronization involves a loss
of kinetic energy by the quark, enough to supply at least the
difference $m_D-m_c \simeq 350$~MeV.
If the quark does not have this kinetic energy to lose, it may be
trapped in the plasma.
It could emerge later if it gains enough energy through diffusion;
otherwise it is trapped in the plasma until the plasma finally
evaporates.
Of course, it must hadronize at least once.
Whenever the quark/meson is in a region of hadron gas, it is liable to
scatter off the hadrons, mostly pions, and thus to undergo yet more
drag and diffusion.

The situation is not too different if there is no real phase transition.
The high- and low-temperature regimes are still distinguished by very
different entropy densities.  The ``mixed phase'' in this case would be
a mixture of regions with temperatures slightly above and slightly
below the crossover temperature at which the entropy changes steeply.
The notion of boundaries between droplets of the pure phases should
still be valid, as should the estimates of $\tau_p$~and~$\tau_h$.

For simulation of the mixed phase we adopt the cascade hydrodynamics
model \cite{Bertsch} of Bertsch {\em et al.}
In this model, when the temperature reaches $T^*$ the plasma simply
breaks up into droplets.
As the system continues its longitudinal expansion the droplets become
ever more rarefied, all the while radiating and reabsorbing pions.
Finally the droplets all evaporate, completing the phase transition.
We have supplemented the cascade code with the routines needed to
follow the progress of a charmed quark/meson through the
system.

In the next section we describe the various assumptions of our model
and how they are implemented in the simulation program \cite{thesis}.
Proceeding step-by-step through the collision process, we discuss the
initial conditions given the plasma and the charmed quark diffusing
therein; the Langevin process by which this diffusion takes place; and the
cascade hydrodynamics by which the plasma droplets and pions are
simulated in the mixed phase\footnote{
This is essentially unchanged from the algorithm originally used by
Bertsch {\em et al.\/} \cite{Bertsch}}.
We then present the model which is central to the physics presented
here, that of charm hadronization and re-ionization \cite{bqs_mixed},
and finally discuss the $D\pi$ scattering which occurs when the charmed
quark is outside the plasma.

In Section 3 we present our results, which are mainly predictions for
the final $p_\perp$ of a charmed meson born of a $c$ quark created with
initial $p_\perp=p_{\perp0}$.
We plot this quantity for a range of assumptions concerning the
free parameters of the simulation.
Our main {\em qualitative\/} result may be simply stated: 
$D$ mesons do provide a rough thermometer of the phase transition.
This is because a large percentage of the charmed quarks are trapped
in the plasma until it breaks up;
moreover, those which escape the plasma droplets early will experience
many collisions in the pion gas which has not yet rarefied.
Unfortunately, the temperature scale of the thermometer 
contains uncertainties due to the dynamical assumptions we make,
chiefly the initial size of the plasma droplets.
The measurement of temperature is also degraded
by transverse flow of the droplets and pions.

\section{Ingredients of the simulation}

\subsection{Initial conditions}

The initial conditions we specify are of two classes.
One describes the cylinder of plasma at proper time $\tau_0$, while the
other gives the initial momentum and location of each charmed quark.

The plasma is assumed to occupy a cylinder of fixed
cross section $R$, equal to the radius of the colliding nuclei.
We consider only central $UU$ collisions, and hence
$R\simeq 7$~fm.
The initial time $\tau_0$ is the time at which the plasma is taken
to be fully formed.
We can estimate it from the flux tube model \cite{flux_tube} according to
\cite{KMS} $\tau_0\simeq (1~\hbox{fm})\cdot A^{-1/6}\simeq 0.5$~fm.

We fix the initial temperature in the usual way
\cite{Satz} by setting the eventual pion multiplicity
$dN_\pi/dy$.
This is related to the entropy density at freezeout by
$dS/dy=3.6\,dN_\pi/dy$.
The cascade simulation predicts an entropy increase of 20--30\%
during the phase transition \cite{Bertsch}; if we take this into
account, while neglecting
the entropy created through dissipation at other
stages of the expansion, then the entropy density $dS/dy$
at $\tau_0$ is $(1.2)^{-1}$ times that at freezeout.
Setting
\begin{equation}
\frac{dS}{dy}=\left(8+6N_f\frac78\right)\frac{4\pi^2}{45}T^3
\cdot \pi R^2\tau  
\end{equation}
in the quark-gluon plasma, we determine the temperature $T_0$
at $\tau_0$.
(For simplicity we take $N_f=2$.)
We present results for an initial temperature of 300~MeV, corresponding
to $dN/dy=1400$. 


The hard collision of the incident nuclei takes place at $t=z=0$,
and hence the charmed quarks are created on trajectories emanating from
this surface and thus satisfying $z=v_\ell t$ for some $v_\ell$.
We follow one quark at a time,
and so the longitudinal-boost invariance of the fluid allows us to work
in the frame where $v_\ell=0$.
By choosing the $x$ axis along the quark's initial position we end up,
with no loss of generality, with\footnote{
Actually for reasons of economy we put two charmed quarks into
each event.
We make sure that they have no effect on each other by giving them different
initial values of $x_0$, with opposite sign, and the same ${\bf p}_0$.}
${\bf x}_0=(x_0,0,0)$ and ${\bf p}_0=(p_0^x,p_0^y,0)$.

The radius $x_0$ at which the heavy quark is created is taken from a
distribution proportional to the number of nucleon--nucleon collisions
taking place at that radius,
\begin{equation}
p(x_0)\,dx_0 \propto \left(R^2-x_0^2\right) 2\pi x_0\,dx_0\ .
\end{equation}
Azimuthal symmetry of the parton--parton collision gives the direction
of ${\bf p}_0$ in the $xy$ plane a uniform distribution.
The magnitude $p_0$ is an input parameter.

Charmed quarks are of course created in pairs.
We neglect entirely the interaction between the quark and antiquark,
and any influence they may have on each other as they diffuse through
the plasma and mixed phases.
The reason is that the two are created with opposite transverse
momenta (and different rapidities) which carry them apart quickly.
Since $\tau_p$ is (under our assumptions) at least 1.7~fm$/c$,
the quarks can be 3~fm apart
by the onset of the phase transition; since the droplets in the
mixed phase have initial radii of 1--2~fm (see below),
the quark and antiquark will
generally find themselves in different droplets.\footnote{
The rare case where the $c\bar c$ pair are found in the same
droplet might make a significant contribution to $J/\psi$
production.}
Their interaction will therefore be indirect at best.
We have not studied correlations in the {\em directions} of the produced
$D$ pair \cite{Hwa}, which may offer another diagnostic probe
of the phase transition.

\subsection{Diffusion before the phase transition}

From $\tau=m_c^{-1}$ to $\tau=\tau_0$ the interaction region is
in a rapidly changing state of particle creation and equilibration.
Since this period is very short, it doesn't make much difference how
we model it.
For simplicity, we proceed as if it were an equilibrium
plasma at $T=T_0$.
We allow the charmed quark to diffuse in this plasma in the same
way as it does after $\tau_0$, which we now describe.

From $\tau_0$ to $\tau_p$ the charmed quark is in a pure plasma.
We model its diffusion by a non-relativistic Langevin equation,
\begin{equation}
\frac{d{\bf p}}{dt} = -\gamma(T) {\bf p} +\mbox{\boldmath$\eta$}\ ,
\label{Langevin}
\end{equation}
where {\boldmath$\eta$} is a Gaussian noise variable, normalized such that
\begin{equation}
\left\langle\eta_i(t)\eta_j(t')\right\rangle=\alpha(T)
\delta_{ij}\delta(t-t')\ .
\end{equation}
Both the drag coefficient $\gamma$ and the momentum-space diffusion
coefficient $\alpha$ depend on the local temperature.
We take $\gamma$ from the result of \cite{bqs_drag}, which we
parameterize as\footnote{We here correct an error in
\cite{bqs_mixed}.}
\begin{equation}
\gamma(T)=aT^2,\qquad a=2\times10^{-6}
\,\hbox{fm}^{-1}\,\hbox{MeV}^{-2}\ .
\end{equation}
(We neglect any momentum dependence in $\gamma$.)
$\gamma$ and $\alpha$ are related by the fluctuation-dissipation
relation, which says that in equilibrium
\begin{equation}
\left\langle p_i^2\right\rangle = \frac\alpha{2\gamma}\ .
\label{fluc_diss}
\end{equation}
For temperatures where the quark in equilibrium is non-relativistic, we would
have $\left\langle p_i^2\right\rangle =m_cT$;
in the range $T=$150--250~MeV, however,
$\left\langle p_i^2\right\rangle$ is up to 33\% larger than this.
As a relativistic correction, we calculate $\alpha$ from
(\ref{fluc_diss}) with $\left\langle p_i^2\right\rangle =1.33\,m_cT$.

We apply (\ref{Langevin}), which is not Lorentz-invariant,
in the rest frame of the plasma surrounding
the quark.
In the Bjorken scaling expansion we assume, this is a frame moving 
longitudinally with
$v_z=z/t$, where $z$ and $t$ are the coordinates of the quark.
The local temperature is given by $T=T_0(\tau_0/\tau)^{1/3}$.

\subsection{Simulation of the mixed phase}

The cascade hydrodynamics model \cite{Bertsch} of the mixed phase
assumes that at proper time $\tau_p$, the beginning of the phase
transition, the plasma breaks up into droplets.
These droplets subsequently emit and absorb pions.
The droplets' interior is always at the transition temperature
$T^*$ and thus their energy density is fixed at
\begin{equation}
\varepsilon=\left(8+6N_f\frac78\right)\frac{\pi^2}{15}T^{*4}+B\ ,
\end{equation}
where $B$ is the bag constant.
We worked with two values for the transition temperature,
$T^*=150$~MeV and $T^*=200$~MeV, with $B$ adjusted accordingly.
Table \ref{parameter_table}
shows the values of various quantities for the two cases.

A droplet with radius $r_d$ has mass equal to
\begin{equation}
M_d=\frac43\pi r_d^3\varepsilon\ .
\end{equation}
The initial value of $r_d$ determines how long it takes the droplets
to evaporate.
If the transition were adiabatic, the lifetime
of the mixed phase would be determined by entropy conservation, which 
makes the ratio of times equal to the ratio of entropy
densities,
\begin{equation}
\frac{\tau_h}{\tau_p}=\frac{\sigma_p}{\sigma_h}\simeq10\ .
\end{equation}
In the cascade model, however, the expansion is {\em not} adiabatic
and the duration of the transition, that is, the time needed
for the droplets to evaporate, is strongly dependent on the droplets'
initial radius.
Thus larger initial radii give a longer-lived mixed phase.\footnote{
We find that for $r_0=1.0,$ 1.5, and 2.0~fm the duration of the
phase transition turns out to be 12, 22, and 32~fm/$c$, respectively,
independent of the transition temperature.}
We experimented with different initial values for $r_d$ ranging from
1~fm to 2~fm.
As each droplet emits and absorbs pions its mass changes, and $r_d$
changes accordingly.

\begin{table}
\begin{tabular}{lcc|c|c}
Transition temperature&$T^*$&&150 MeV&200 MeV\\ \hline
Beginning time of transition&$\tau_p$&(fm)&
4.0&1.7\\
Bag constant&$B^{1/4}$&(MeV)&208&278\\
Plasma energy density&$\varepsilon$&(GeV/fm$^3$)&1.0&3.2\\
Thermal pion energy&$E_{\pi}^T$&(MeV)&480&610\\
Thermal pion momentum&$p_{\pi}^T$&(MeV)&520&680\\
Thermal $D$ momentum&$p_D^T$&(MeV)&1010&1200
\end{tabular}\smallskip
\caption{\label{parameter_table}
Physical quantities for the two transition temperatures
considered.
$p_{\pi}^T$ and $p_D^T$ are root-mean-square momenta, 
$p^T\equiv\protect\sqrt{\langle{\bf p}^2\rangle}$, 
while $E^T\equiv\langle E\rangle$.
}
\end{table}

The program actually steps through the time coordinate $t$ rather than
the proper time $\tau$.
At each time $t$ we determine the $z$ coordinates of the two points on
the hyperbola $\tau=\tau_p$, where $\tau\equiv\sqrt{z^2-t^2}$:
This is where the phase transition is currently occurring, and thus
where droplet creation should take place.
The droplets are uniformly distributed across the cylinder, and their
number in the $z$-slice at $\tau_p$ is fixed to conserve the entropy
of the plasma at the breakup.
A droplet's initial momentum is thermally distributed in the locally
comoving rest frame, which has rapidity $y=\tanh^{-1}(z/t)$.

From the time of its formation, each droplet interacts with the pion
gas.
By detailed balance, the pion emission rate is equal to the
absorption rate in equilibrium \cite{Weisskopf}.
The latter is determined by assuming that a droplet absorbs all pions
that approach within a distance $d=1$~fm from its surface.
The absorption rate in equilibrium is thus 1/4 of the pion flux times
the area of the extended droplet surface,
\begin{equation}
W=\frac14n_\pi^{\rm eq}v_\pi \cdot 4\pi (r_d+d)^2\ .
\end{equation}
We take
\begin{equation}
n_\pi^{\rm eq}=3\frac{16}{\pi^2}\zeta(3)T^3
\end{equation}
and $v_\pi=1$, as appropriate for ultrarelativistic pions.
Each emitted pion is created at a random place on the droplet's
surface, with a thermally distributed momentum (in the droplet's rest
frame) directed outward.
The emission process conserves energy and momentum exactly
through the recoil and shrinkage of the rigid droplet.
Pion absorption is simpler, consisting of the disappearance of any
pion which ventures within a distance $d$ of a droplet's surface (in
its rest frame), with its four-momentum taken up by the droplet's
recoil and growth.

If the droplet contains less rest energy than that required to make
four thermal pions (see Table \ref{parameter_table}),
it falls apart altogether into $n$ thermally distributed pions.
$n$ might in fact be less than four, because if $M_d/m_\pi<4$ then
$n$ is given by the integer part of this ratio.
As in \cite{Bertsch}, we adopt the simple model of first giving each
pion a random, thermally-distributed momentum; the momenta are then
shifted and rescaled to conserve the momentum and energy of the droplet.

$\pi\pi$ scattering is taken to be isotropic, with an
isospin- and energy-depend\-ent cross section of about
30~mb.\footnote{This is the weighted average at $T=200$~MeV.
See \cite{Bertsch} for details.}
Droplet--droplet scattering is neglected entirely, which is
reasonable after the earliest times because
dilution by the longitudinal expansion is rapid.

\subsection{Charm hadronization and re-ionization}

At the beginning of the phase transition the charmed quark is inside
a droplet.
(We restart the simulation if the random creation of droplets leaves
the quark outside.)
It continues to undergo diffusion in the plasma as it did before the
transition time.
The Langevin equation (\ref{Langevin}) is applied in the droplet's rest
frame via the appropriate Lorentz transformation, and every change in
the quark's momentum is balanced by recoil of the  droplet
(and a change in its mass to conserve energy).
Sooner or later, the quark strikes the wall of the droplet.

Our model of hadronization is based on a picture of
a quark incident on a static bag boundary \cite{Danos,Matsui,Eisenberg}.
Upon striking the wall the quark stretches a flux tube \cite{Matsui}
outward which we assume to be always radially oriented.
The tube possesses a tension (energy density)
$\sigma=0.16~\hbox{GeV}^2$ in the droplet rest frame, and we take
this to be the only force acting on the quark during this
period.
Thus we integrate the equation
\begin{equation}
\frac{d{\bf p}}{dt}=-\sigma\hat{\bf r}
\end{equation}
in the droplet frame until either the quark reenters the droplet
or the flux tube breaks.\footnote{The droplet also reacts to the
tension of the tube.}

The flux tube fissions at a rate $d\Gamma/d\ell$ per unit length.
Various estimates for this quantity are 0.3~fm$^{-2}$ from the flux tube
model \cite{Matsui} and anywhere from 0.5 to 2.6~fm$^{-2}$ from
comparisons of string fragmentation models with experiment \cite{string}.
We adopt two values for $d\Gamma/d\ell$ at the extremes of the
range, 0.5 and 2.5~fm$^{-2}$.

The tube breaks at a point $\cal P$ some distance $\ell$ from its base.
The segment from $\cal P$ to the droplet is reabsorbed in the droplet;
the remainder of the tube snaps back into the quark and turns it into a meson.
The $D$ emerges with energy (in the droplet rest frame)
$E_D=E_c+\sigma(L-\ell)$, where $L$ is the total length of the tube.
Clearly $\sigma(L-\ell)$ cannot be less than $m_D-E_c$, or there will not
be enough energy for hadronization at all.\footnote{We neglect the
possibility of fragmentation of the quark into a $D^*$ or a $D\pi$
pair, because of the sharply reduced phase space for such
fragmentation.}

A quark which strikes the droplet wall with
energy $E_c<m_D$ would be unable to hadronize if the
droplet were static.
Its reflection from the droplet wall, however, still involves
stretching a flux tube and this process can take 1.5~fm/$c$ in the droplet rest frame.
During this time, the droplet emits and absorbs pions and thus
its recoil might stretch the flux tube further.\footnote{A change
in the droplet's radius also causes the flux tube to change
its length, conserving energy via the droplet's mass.}
This effect, which of course applies as well to situations
with $E_c>m_D$, sometimes enables even quarks which are initially
below threshold to hadronize.
Again, the flux tube breaks at a random point, with the constraint
$\sigma(L-\ell)>m_D-E_c$ in the droplet's new rest frame.

When the tube breaks, the resulting $D$ meson emerges immediately with
its momentum in the same direction as that of the quark.
Four-momentum is conserved via recoil of the droplet.
In the droplet's initial rest frame, the kinetic energy of recoil
comes from the attached segment of the flux tube
and from a change in its rest mass,
\begin{equation}
M_d^{\prime2}+(\Delta p)^2=(M_d+\sigma\ell)^2\ ,
\label{recoil}
\end{equation}
where $\Delta p\equiv p_c-p_D$ is the momentum transfer from the
quark/meson.

It is possible for the $D$-meson to be created with momentum directed
towards the droplet, and even for it to collide with the droplet
in the next time step.
At that point the quark is reabsorbed in an ordinary ionization
process (see below), which means it has been effectively reflected
back into the droplet.

We neglect entirely an additional possibility for hadronization,
namely, that the $c$ quark might encounter a light antiquark as it
nears the droplet surface.
It could thus hadronize without forming a flux tube, and possibly
even gain energy in the process.
We calculate the probability of this to be small, as follows.
The density of light antiquarks at $T=200$~MeV is $n_q=1~{\rm
fm}^{-3}$, whence the flux of outward-going antiquarks at the droplet
surface is $j_q=n_q/4$.
An antiquark must appear within a range $R\simeq0.5$~fm of the
$c$ quark during the reflection process which takes $\Delta t\simeq
1.5$~fm/$c$.
The relative probability of forming a color singlet is 1/9.
Thus the probability of hadronization through light-quark binding is
only
\begin{equation}
P=j_q\cdot(\pi R^2)\cdot\Delta t\cdot\frac19\simeq 3\%.
\end{equation}

The only remaining possibility for hadronization is that the droplet
breaks up around the $c$ quark, which happens
when the mass of the droplet falls below the energy needed to make
four thermal pions {\em plus the energy needed for
hadronization.}
In our scenario, the charmed quark picks up matter which was at rest in
the droplet rest frame, and thus its three-momentum is conserved,
$p_D=p_c$, while the energy required for creation of the $D$ is
supplied by the droplet.
Having given this energy to the meson, the droplet then breaks
up according to the rule given above for ordinary droplet
breakup.

If the $c$ quark is inside a flux tube at the time the droplet breaks up,
the flux tube's energy is merely added to that of the droplet, and
the remainder of the breakup is handled as above.

The inverse of hadronization, ionization, occurs when a $D$ meson
strikes a droplet, and is easily dealt with.
We assume that the light quark is simply stripped off the meson,
leaving a $c$ quark with the same velocity $v_c=v_D$ (a
frame-independent statement).
This means that the quark/meson loses energy since $E_c=(m_c/m_D)E_D$.
As noted in \cite{bqs_mixed}, a $D$ meson which undergoes ionization
and re-hadronization in quick succession would lose energy in the
amount
\begin{eqnarray}
\nonumber
\Delta E&=&\Delta E_{\rm ion}+\Delta E_{\rm had} =
E_D\left(1-\frac{m_c}{m_D}\right)+0 \\
&\geq&m_D\left(1-\frac{m_c}{m_D}\right)
\simeq 370~{\rm MeV}\ .
\label{DeltaE}
\end{eqnarray}
A quark with less kinetic energy than this will get trapped in the
droplet.
Of course, our simulation actually follows in detail the diffusion of
the $c$ quark inside the droplet so the estimate (\ref{DeltaE}) is
only a rough guide.

\subsection{Scattering in the hadron gas}


Whenever the charmed quark is outside a droplet it moves as
a $D$ meson.
It can then scatter off the pions which have been emitted from the
droplets.
We use a simple model of isotropic, energy-independent, and
charge-independent scattering, with total cross-section
$\sigma_{D\pi}=9$~mb.
We arrive at this number via the additive quark model, as follows.
Beginning with the (high-energy) $pp$ cross section,
$\sigma_{pp}=40$~mb, we estimate that the cross section for light
quark scattering is
$\sigma_{qq}=\sigma_{pp}/9\simeq 4$~mb.
Similarly we begin with the $\psi N$ cross section \cite{psiN},
$\sigma_{\psi N}=2$~mb, and estimate that the cross section for charmed
quarks on light quarks is $\sigma_{cq}\simeq 0.3\ {\rm mb}\ll\sigma_{qq}$.
Thus we arrive at $\sigma_{D\pi}=2\sigma_{qq}+2\sigma_{cq}\simeq
9$~mb.
We ignore resonant scattering via the $D^*$ since the resonances lie
at $p=39$--44~MeV/$c$ in the center-of-momentum frame, which is
far below the typical thermal momenta (see Table \ref{parameter_table}).

(This procedure gives good results for $K\pi$ scattering.
Here one begins with $\sigma_{\phi N}=13.8$~mb \cite{phiN} and reaches
$\sigma_{K\pi}\simeq13$~mb, which is not far from the
values from phase shift analysis \cite{Kpi} 8~mb$<\sigma_{K\pi}<13$~mb
in the energy range 1~GeV$<\sqrt{s}<1.5$~GeV, above the $K^*$ resonance.)

\section{Numerical results}

The aim of our calculation is a prediction for the $p_\perp$ distribution
of $D$ mesons created in heavy ion collisions.
The result naturally depends on the distribution one assumes
for the initial momentum of the $c$ quark.
Rather than tie ourselves to specific model predictions for the latter
\cite{charm_production},
we present our results as plots of the RMS $p_\perp^D$ for
{\em given} initial quark momentum $p_{\perp0}^c$ in the range
between 1~GeV and 2~GeV.

Our main result is apparent in the last figure,
which shows that $p_\perp^D$ can be a thermometer of the mixed
phase.
We begin, however, with a detailed analysis of the numerical results
for $T^*= 150$~MeV.
\begin{figure}\vspace{-3cm}
\centerline{\psfig{figure=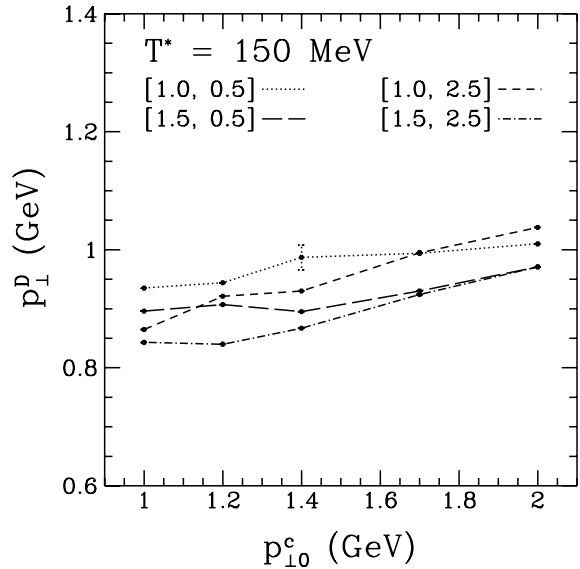,height=8cm}}
\caption{\label{T150final}
RMS transverse momentum of $D$ meson vs.~initial transverse momentum
of $c$ quark, for transition temperature $T^*=150$~MeV.
The four curves correspond to different values of $r_0$ and $d\Gamma/d\ell$
(in fm and fm$^{-2}$, respectively):
[1.0, 0.5] (dotted), [1.0, 2.5] (short dashes), [1.5, 0.5] (long dashes),
[1.5, 2.5] (dashed-dotted).
A typical statistical error bar is shown.
Results with $r_0=2.0$~fm (not shown) are close to, and just below, the
corresponding points for $r_0=1.5$~fm.}
\end{figure}
In Fig.~\ref{T150final}
 we show the calculated $p_\perp^D$ for two values of
the initial droplet radius
$r_0$ and of the flux-tube fission rate $d\Gamma/d\ell$, as discussed above.
We note two features immediately:
(1) $p_\perp^D$ varies but weakly with $p_{\perp0}^c$, showing the
effects of drag and thermalization;\footnote{The dependence will be
even weaker for $p_{\perp0}^c<1$~GeV because thermalization will wipe
out any memory of the initial momentum.}
(2) $p_\perp^D$ lies generally above its thermal value
$\sqrt{\frac23}p^T_D=820$~MeV (see Table 1), showing the effects of 
transverse hydrodynamic expansion.
Moreover, for any given $p_{\perp0}^c$, the width of the distribution of
$p_\perp^D$ (not shown)
is consistent with a thermal distribution at $T=150$~MeV.

In comparing the four data sets in Fig.~\ref{T150final}, 
we find that raising either
$r_0$ or $d\Gamma/d\ell$ causes a drop in $p_\perp^D$ (for $p_{\perp0}^c<
1700$~MeV).
We explain this by selecting two populations of $D$ mesons:
those which emerge from flux-tube fission and then escape the
system---{\em fragmentation mesons\/}---and
those which are trapped within their original droplets until the latter
evaporate---{\em breakup mesons\/}\@.
\begin{figure}\vspace{-3cm}
\centerline{\psfig{figure=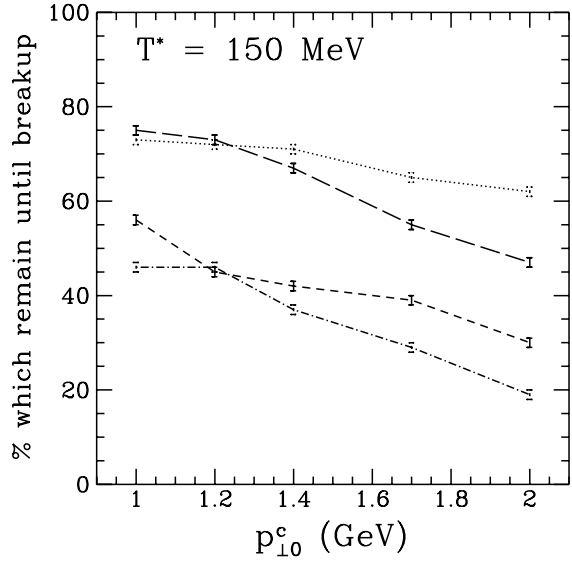,height=8cm}}
\caption{\label{T150trapped}
The proportion of $c$ quarks which are trapped inside their
original droplets until
their breakup, as a function of initial transverse momentum.
The four curves are for different values of $r_0$ and $d\Gamma/d\ell$
as in Fig.~\protect\ref{T150final}.}
\end{figure}
Fig.~\ref{T150trapped}
shows the proportion of breakup mesons for the four data
sets.
This fraction drops with increasing $p_{\perp0}^c$ and with increasing
$d\Gamma/d\ell$, because each of these helps the charmed quark break
a flux tube and escape;
the fraction also drops with increasing $r_0$ because a larger droplet
has a longer lifetime and hence affords greater opportunity for escape
before breakup.
Note that the fraction of breakup mesons lies between 20\% and 75\%,
so the thermalization of charmed quarks in the plasma is a very
important effect.

\begin{figure}\vspace{-1cm}
\centerline{\psfig{figure=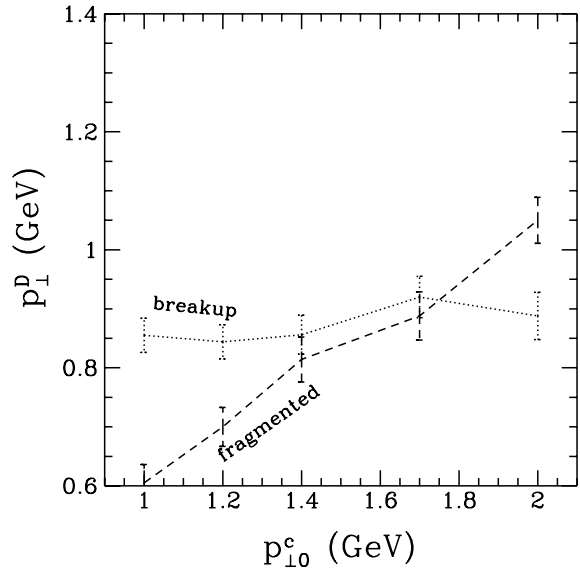,height=8cm}}
\caption{\label{T150trapped2}
Comparison of two populations of $D$ mesons.
The RMS $p_\perp$ of $D$ mesons is plotted against the initial
$p_\perp$ of the $c$ quarks, for breakup mesons 
(dotted curve) and for fragmentation mesons (dashed curve).
Here $T^*=150$~MeV, $r_0=1$~fm, $d\Gamma/d\ell=2.5$~fm$^{-2}$.}
\end{figure}
In Fig.~\ref{T150trapped2}
we see that breakup mesons are strongly thermalized, that is,
their $p_\perp$ is independent of $p_{\perp0}^c$.
(Fig.~\ref{T150trapped2}
shows results of a simulation with $D\pi$ scattering
turned off, so that hydrodynamic effects are weak and $p_\perp^D$ is
close to its thermal value.)
The $p_\perp$ of the fragmentation mesons, however, does depend on
$p_{\perp0}^c$ and for low values of $p_{\perp0}^c$ it in fact
lies {\em below} the thermal value; this is due to our static model
for the droplet wall, which does not include thermal
fluctuations in its position and velocity caused by coupling to the
plasma inside.
Thus for low values of $p_{\perp0}^c$ the proportion of fragmentation
mesons affects $p_\perp^D$ strongly.
Having noted the dependence of this proportion on the parameters
$r_0$ and $d\Gamma/d\ell$, we now understand the trends
in Fig.~\ref{T150final}.

\begin{figure}\vspace{-3cm}
\centerline{\psfig{figure=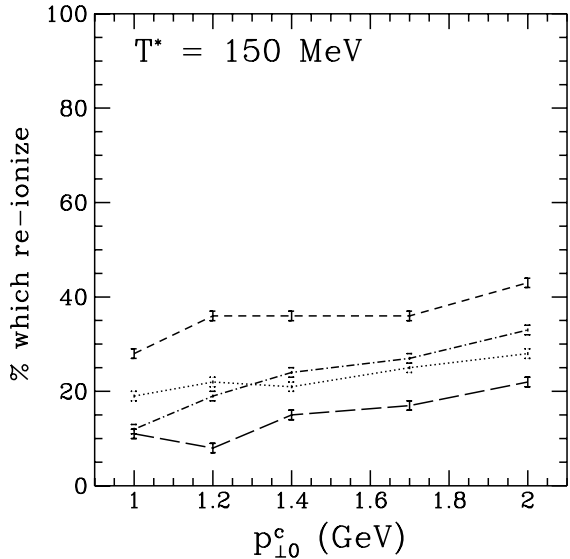,height=8cm}}
\caption{\label{T150reion}
The proportion of $c$ quarks which undergo re-ionization,
as a function of initial transverse momentum.
The four curves are for different values of $r_0$ and $d\Gamma/d\ell$
as in Fig.~\protect\ref{T150final}.}
\end{figure}
These trends are strengthened by the phenomenon of re-ionization, that is,
reabsorption
by a droplet after initial hadronization.
The proportion of quarks which undergo re-ionization
is shown in Fig.~\ref{T150reion}.
This process sharply lowers the momentum
of $c$ quarks.
The lower the quark momentum, the less the chances of fragmentation
and the greater the probability that the eventual meson will emerge
only at breakup of the droplet.
This process thus converts fragmentation mesons into 
a second generation of breakup mesons.
(The latter are {\em not} represented in Figs.~\ref{T150trapped} and
\ref{T150trapped2}.)
The dependence on the parameters is explained as follows.
Raising $d\Gamma/d\ell$  will lead to earlier release
of the $D$ meson, which puts it into an environment with higher
droplet density, which will increase the chances of recapture;
lowering $r_0$ will increase the number of droplets as
$r_0^{-3}$, which again will make re-ionization more likely
despite the decreased cross section
for capture.
 
\begin{figure}\vspace{-3cm}
\centerline{\psfig{figure=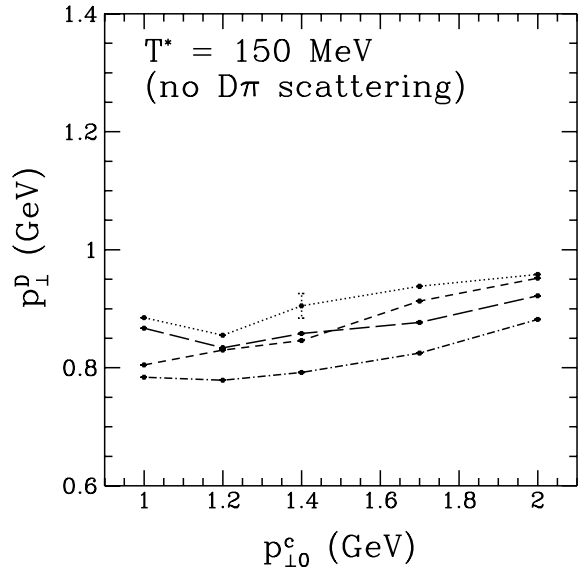,height=8cm}}
\caption{\label{T150noDpi}
RMS transverse momentum of $D$ meson vs.~initial transverse momentum
of $c$ quark, with $D\pi$ scattering suppressed.
The four curves are for different values of $r_0$ and $d\Gamma/d\ell$
as in Fig.~\protect\ref{T150final}.}
\end{figure}
Finally, to show the importance of $D\pi$ scattering, we show results
(Fig.~\ref{T150noDpi}) of simulations in which it is omitted.
Comparison of Fig.~\ref{T150noDpi} with Fig.~\ref{T150final}
 shows that $D\pi$ scattering causes
a systematic increase in $p_\perp^D$ of about 60~MeV.
We summarize our findings at $T^*=150$~MeV as follows:
 (1) Break-up mesons emerge from
the droplets with thermal momenta; (2) fragmentation mesons emerge with {\em
lower than\/} thermal momenta; (3) $D\pi$ scattering adds momentum so
that the resulting $p_\perp^D$ is in the neighborhood of its thermal
value.

More interesting than this detailed analysis, however, is a comparison
to the case of $T^*=200$~MeV, to which we now proceed.
\begin{figure}\vspace{-3cm}
\centerline{\psfig{figure=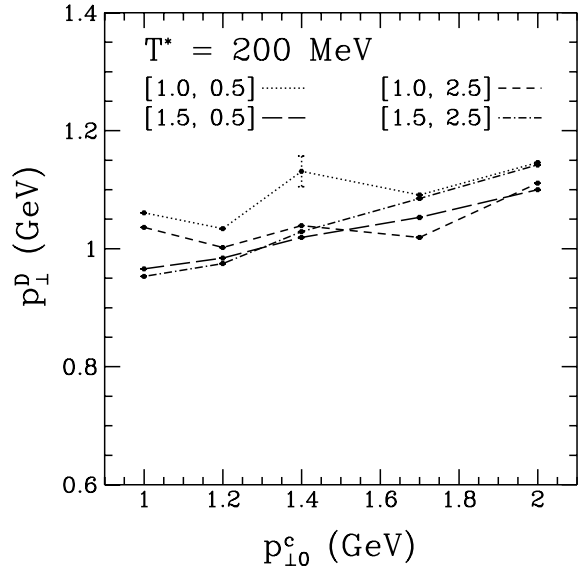,height=8cm}}
\caption{\label{T200final}
As in Fig.~\protect\ref{T150final}, but for $T^*=200$~MeV.}
\end{figure}
Fig.~\ref{T200final} shows that the RMS
$p_\perp^D$ here behaves much the same as for
$T^*=150$~MeV, but with a systematic shift upward of 120~MeV
for given $r_0$ and $d\Gamma/d\ell$.
(The thermal value of this quantity is 980~MeV, an increase
of 160~MeV from $T^*=150$~MeV.)
\begin{figure}\vspace{-1cm}
\centerline{\psfig{figure=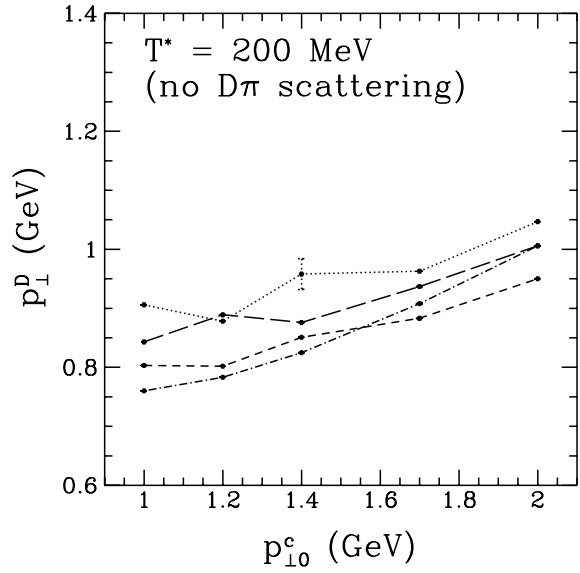,height=8cm}}
\caption{\label{T200noDpi}
As in Fig.~\protect\ref{T200final} ($T^*=200$~MeV), but with $D\pi$ scattering 
suppressed.}
\end{figure}
As may be seen in Fig.~\ref{T200noDpi}, the bulk of this effect is due to
$D\pi$ scattering:
Comparison of Fig.~\ref{T200noDpi} with Fig.~\ref{T150noDpi} shows, in the 
absence of $D\pi$ scattering,
almost no difference between the two temperatures.
\begin{figure}\vspace{-3cm}
\centerline{\psfig{figure=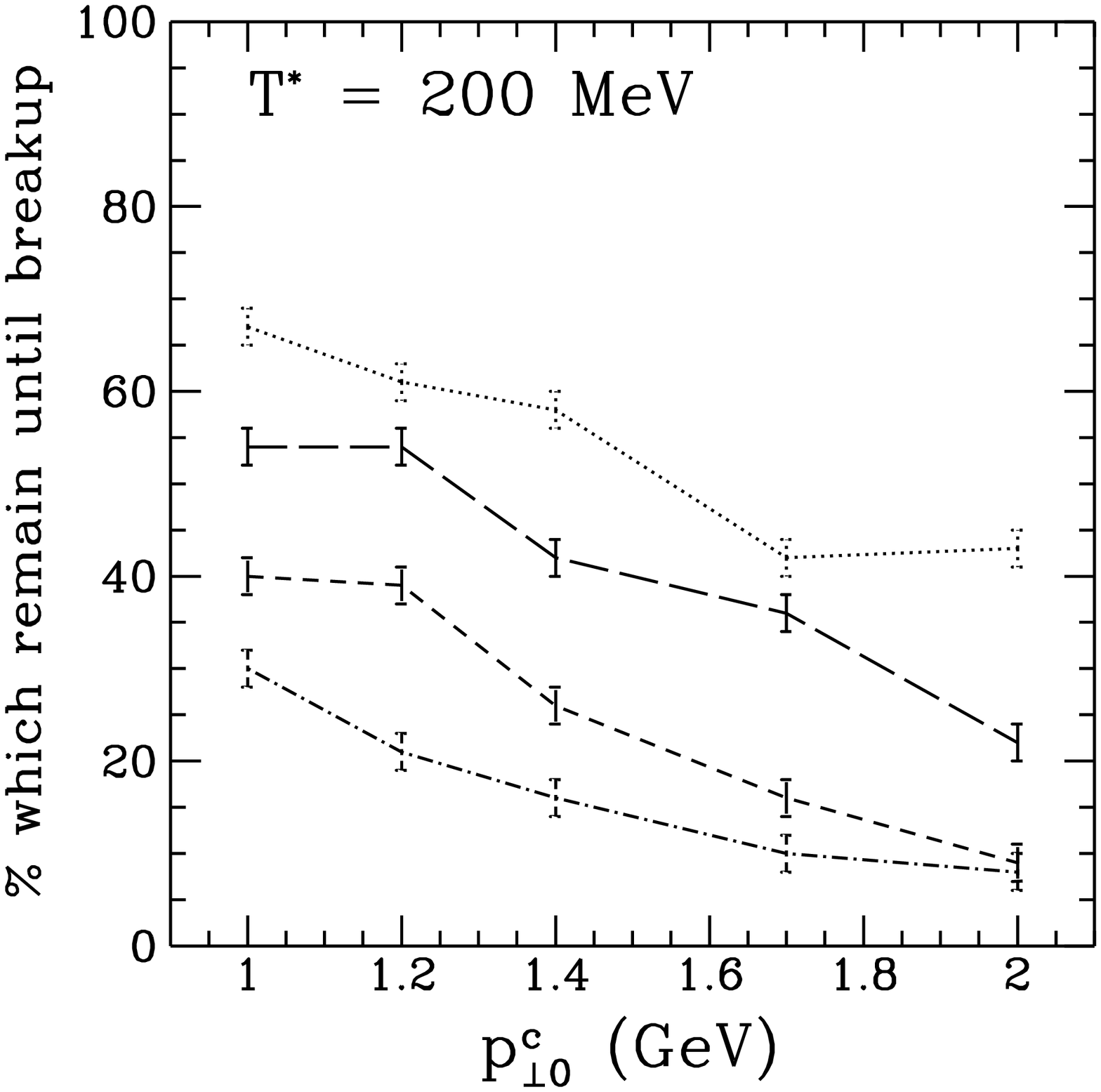,height=8cm}}
\caption{\label{T200trapped}
As in Fig.~\protect\ref{T150trapped}, but at $T^*=200$~MeV.} 
\end{figure}
Examination of the counterpart to Fig.~\ref{T150trapped}
shows (see Fig.~\ref{T200trapped}) that the
number of breakup mesons has decreased substantially, mainly because
thermal motion of $c$ quarks in the droplets is more effective at
pushing them over fragmentation thresholds.
Having seen that breakup mesons are thermalized while fragmentation mesons
are not,
we understand why the $D$ mesons, upon emerging from the droplets,
are no hotter in spite of the increase in the ambient temperature.
It is only the interaction with the hot pion gas that heats the
$D$ mesons to thermal momenta and restores their thermometric value.

An experimental measurement of the $p_\perp$ distribution of $D$ mesons
will reflect a folding of one of the curves in Fig.~\ref{T150final}
(or~\ref{T200final}) with
a distribution for $p_{\perp0}^c$.
As a source of uncertainty, the weak $p_{\perp0}^c$ dependence is not
very troublesome.
More significant are the differences among the curves with different
parameters $r_0$ and $d\Gamma/d\ell$.
\begin{figure}\vspace{-1cm}
\centerline{\psfig{figure=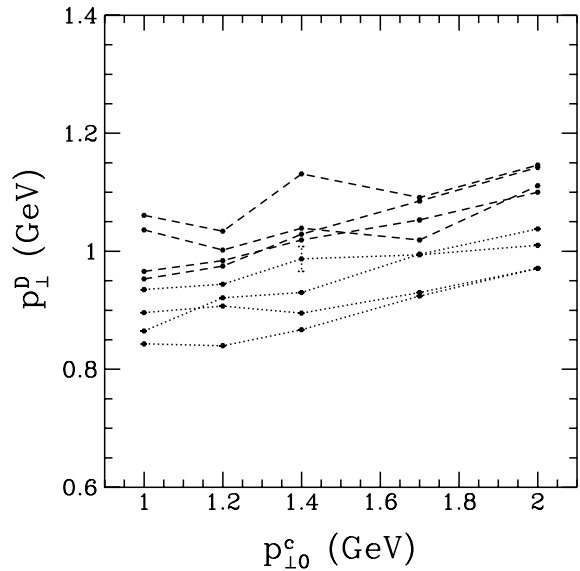,height=8cm}}
\caption{\label{bothT}
Comparison of results for $T^*=150$~MeV (dotted curves, same data as in
Fig.~\protect\ref{T150final}) with results for $T^*=200$~MeV (dashed
curves, same data as in Fig.~\protect\ref{T200final}).} 
\end{figure}
Nevertheless,
as seen in Fig.~9, there is almost
no overlap between the range of $p_\perp^D$ for $T^*=150$~MeV and
that for $T^*=200$~MeV.
A measurement of $p_\perp^D$ will still provide a rough thermometer
of the transition temperature, one that can become finer as the assumptions are
narrowed down.

Of the parameters we vary, we find that the initial droplet radius
affects the $p_\perp^D$ most strongly; unfortunately, this parameter
is the one which is furthest from quantitative understanding.
Improvement of our results must therefore
await a more detailed model of droplet
formation in the phase transition.

\section*{Acknowledgements}
We thank S.~Gavin, G.~Bertsch, and L.~McLerran for discussions concerning
the cascade hydrodynamics simulation code.
This work was supported by a Wolfson Research Award administered by
the Israel Academy of Sciences and Humanities, and by the Basic
Research Fund of Tel Aviv University.
B. S. thanks the European Center for Theoretical Studies in Nuclear
Physics and Related Areas (ECT*) in Trento for its hospitality
while this paper was being written.

\end{document}